# Investigating Cellular Automata Based Network Intrusion Detection System For Fixed Networks (NIDWCA)

P.Kiran Sree [1], I. Ramesh Babu [2]

[1] Research Scholar, Dept of C.S.E, Jawaharlal Nehru Technological University, Hyderabad.
[2] Professor, Dept of C.S.E, Acharya Nagarjuna University, Guntur.
Email: [1] profkiran@yahoo.com, [2] rinampudi@hotmail.com

**Abstract**

*Network Intrusion Detection Systems (NIDS) are computer systems which monitor a network with the aim of discerning malicious from benign activity on that network. With the recent growth of the Internet such security limitations are becoming more and more pressing. Most of the current network intrusion detection systems relay on labeled training data. An Unsupervised CA based anomaly detection technique that was trained with unlabelled data is capable of detecting previously unseen attacks. This new approach, based on the Cellular Automata classifier (CAC) with Genetic Algorithms (GA), is used to classify program behavior as normal or intrusive. Parameters and evolution process for CAC with GA are discussed in detail. This implementation considers both temporal and spatial information of network connections in encoding the network connection information into rules in NIDS. Preliminary experiments with KDD Cup data set show that the CAC classifier with Genetic Algorithms can effectively detect intrusive attacks and achieve a low false positive rate. Training a NIDWCA (Network Intrusion Detection with Cellular Automata) classifier takes significantly shorter time than any other conventional techniques.*

## 1. Introduction

Intrusion is the attempted act of using a computer system or computer resources without the requisite privileges, causing willful or incidental damage.

The primary aim of Intrusion Detection System (IDS) is to protect the availability, confidentiality and integrity of critical networked information systems. Intrusion Detection Systems are an important component of defensive measures protecting computer systems and networks from abuse. When an IDS is properly deployed it can provide warnings indicating that a system is under attack. It is critical for intrusion detection in order for the IDS to achieve maximal performance. Intrusion data has traditionally been presented in text form and, given the typical volume, is difficult for security officers to process. More efficient intrusion detection data modeling and management methods will allow one to develop improved IDS that handle large intrusion detection databases with greater reliability and faster retrieval rates. As systems become more complex, there are always exploitable weaknesses due to design and programming errors, or through the use of various socially engineered penetration techniques.

## 2. Cellular Automata (CA)

Cellular automata use localized structures to solve problems in an evolutionary way. CA often demonstrates also significant ability toward self-organization that comes mostly from the localized structure on which they operate. By organization, one means that after some time in the evolutionary process, the system exhibits more or less stable localized structures. This behavior can be found no matter the initial conditions of the automata.

A CA consists of a number of cells organized in the form of a lattice. It evolves in discrete space and time. The next state of a cell depends on its own state and the states of its neighboring cells. In a 3-neighborhood dependency, the next state $q_i(t+1)$ of a cell is assumed to be dependent only on itself and on its two neighbors (left and right) and is denoted as:

where, $q_i(t)$ represents the state of the $i^{th}$ cell at $t^{th}$ instant of time, f is the next state function and referred

$$q_i(t + 1) = f(q_{i-1}(t), q_i(t), q_{i+1}(t)) \quad (1)$$

to as the rule of the automata. The decimal equivalent of the next state function, as introduced by Wolfram, is the rule number of the CA cell.

## 2.1 Fuzzy CA fundamentals:

FCA is a linear array of cells which evolves in time. Each cell of the array assumes a state qi, a rational value in the interval [0,1] (fuzzy states) and changes its state according to a local evolution function on its own state and the states of its two neighbors. The degree to which a cell is in fuzzy states 1 and 0 can be calculated with the membership functions. This gives more accuracy in intrusion. In a FCA, the conventional Boolean functions are AND, OR, NOT.

## 2. 2 Dependency matrix for FCA

Rules defined in Equation. 1 should be represented as a local transition function of FCA cell. That rules (Table 1) are converted into matrix form for easier representation of chromosomes.

$$T = \begin{bmatrix} 1 & 1 & 0 & 0 \\ 1 & 1 & 1 & 0 \\ 0 & 0 & 1 & 1 \\ 0 & 0 & 1 & 1 \end{bmatrix}$$

**Figure 1**. Matrix Representation of Rule

**Example 1:** A 4-cell null boundary hybrid FCA with the following rule <238, 254, 238, 252> (that is, <$(q_i+q_{i+1})$, $(q_{i-1}+q_i+q_{i+1})$, $(q_i+q_{i+1})$, $(q_{i-1}+q_i)$>) applied from left to right, may be characterized by the following dependency matrix.

While moving from one state to other, the dependency matrix indicates on which neighboring cells the state should depend. So cell 254 depends on its state, left neighbor and right neighbor.

Now we represented the transition function in the form of matrix( Figure 1). In the case of complement, FCA we use another vector for representation of chromosome. Rules will be used for inducing dynamism into our project.

## 3. Challenges of any Network Based Intrusion System

The cost of ownership should be lower for an enterprise environment. Network-based IDS must examine all packet headers for signs of malicious and suspicious activity. They have to use live network traffic for real-time attack detection. Therefore, an attacker cannot remove the evidence. They also detect malicious and suspicious attacks as they occur, and so provide faster notification and response. These are not dependent on host operating systems as detection sources. Network-based IDS add valuable data for determining malicious intent.

**Table 1.** Rule of FCA

| Non-complemented rules | | Complemented rules | |
| --- | --- | --- | --- |
| Rule | Next state | Rule | Next state |
| 0 | 0 | 255 | 1 |
| 170 | $q_{i+1}$ | 85 | $\overline{q_i+1}$ |
| 204 | $q_i$ | 51 | $\overline{q_i}$ |
| 238 | $q_i+q_{i+1}$ | 17 | $\overline{q_i+q_{i+1}}$ |
| 240 | $q_{i-1}$ | 15 | $\overline{q_i-1}$ |
| 250 | $q_{i-1}+q_{i+1}$ | 5 | $\overline{q_{i-1}+q_{i+1}}$ |
| 252 | $q_{i-1}+q_i$ | 3 | $\overline{q_{i-1}+q_i}$ |

### 3.1 Genetic Algorithm & CA

The main motivation behind the evolving cellular automata framework is to understand how genetic algorithms evolve cellular automata that perform computational tasks requiring global information processing, since the individual cells in a CA can communicate only locally without the existence of a central control the GA has to evolve CA that exhibit higher-level emergent behavior in order to perform this global information processing.

This framework provides an approach to studying how evolution can create dynamical systems in which the interactions of simple components with local information storage and communication give rise to coordinated global information processing.

## 4. CA frame work for Intrusion

At a theoretical level, cellular automaton intrusion detection models can be analyzed by much the same methods of statistical mechanics as have been used in trying to derive the Navier-Stokes equations for physical fluids from the microscopic dynamics of real molecules. Figure 2 show the Design of CA based Intrusion Detection.

### 4.1 Basic Algorithm

CA Tree Building (Assuming K CA Basins)

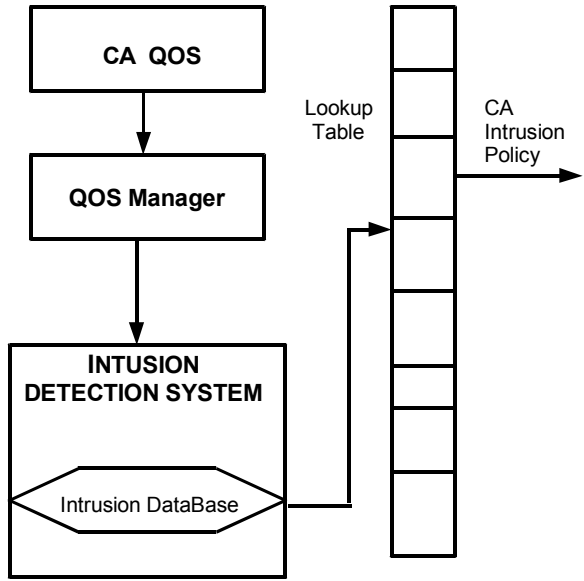

**Figure 2.** Design of CA Intrusion Detection

Input: Intrusion parameters (constraints)Output: CA Based inverted tree Step 0: Start.
Step 1: Generate a CA with *k* number of CA basins
Step 2: Distribute the parameters into k CA basins
Step 3: Evaluate the distribution in each closet basin
Step 4: Calculate the Rq( Malicious Index)
Step 5: Swap the more appropriate features to the bottom leaves of the inverted tree.
Step 6: Stop.

The main feature which is achieved when developing CA Agent systems, if they work, is flexibility, since a CA Agent system can be added , modified and reconstructed, without the need for detailed rewriting of the application.

## 5. Experimental Results

We have applied our algorithm to a set of standard benchmark data, namely the KDD Cup data. Network traffic can be captured using packet-capturing utilities or operating system utilities at the system call level. It is stored as a collection of records in the audit file.

Each data point is described by 41 features. The algorithm was trained with the KDD Cup data set.

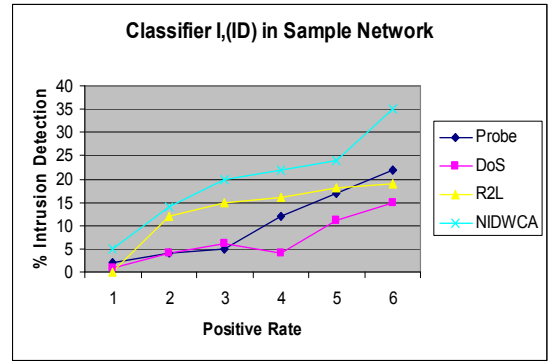

**Figure 3**. Intrusion Detection Vs Positive Rate in Sample Network for Classifier I

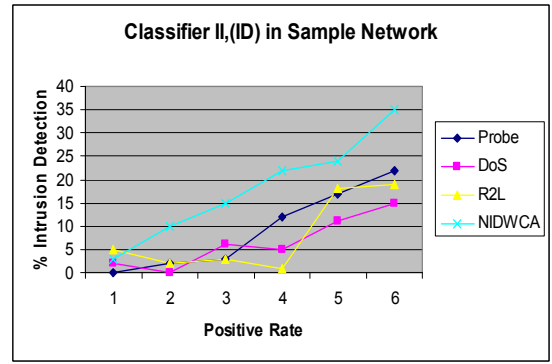

**Figure 4**. Intrusion Detection VS Positive Rate in Sample Network for Classifier II

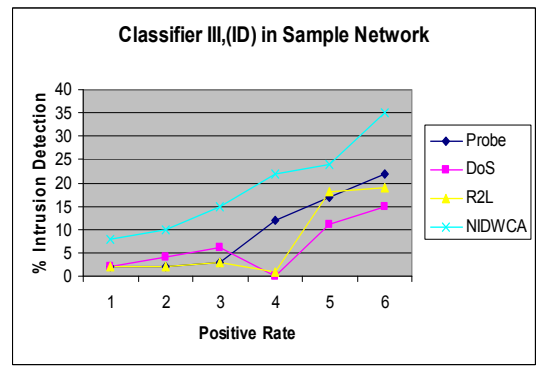

**Figure 5**. Intrusion Detection Vs Positive Rate in Sample Network for Classifier III

We have developed three classifiers for dealing the problem of intrusion. Then we have tested the trained algorithm (4. 3) with the same data set. Figure 3, 4, 5 depicts the findings.

Presently, the work has been conducted only on three sample networks. In other words, the work consists of preliminary findings which offer a base for further expansion and exploration in the study of application of Cellular Automata to Fixed Networks.

## 6. Conclusion

In this paper we have developed NIDWCA an Unsupervised CA based intrusion detection technique strengthened with Genetic Algorithm .It was trained with unlabelled data. The genetic algorithm with CA successfully evolved an individual's model through randomized mutation n and the model generated over training data was successfully able to apply its empirical knowledge to data not seen before. Three types of classifiers were developed to classify intrusion data to recognize whether a system is under attack according to class of networks. The major issue is that the Unsupervised CA trained in a significantly shorter time than any other technique applicable. All these classifiers were found effective in identifying different classes of attacks with low false positive rate.